\documentclass{epl}       
\usepackage{epsfig}                
\def\le{\left}
\def\ri{\right}

\def\ul#1#2{\textstyle{\frac#1#2}}

\def\bnabla{\mbox{\boldmath $\nabla $}}

\shorttitle{Pseudo-Casimir interactions}
\title{Pseudo-Casimir force in confined nematic polymers}
\author{J. Dobnikar\inst{2} \and R. Podgornik\inst{1,2}
\footnote{To whom correspondence should be
addressed. rudolf.podgornik@fiz.uni-lj.si}}
\institute{
	\inst{1} Department of Physics, Faculty of Mathemathics and
                 Physics, University of Ljubljana, Jadranska 19,
                 1000 Ljubljana, Slovenia\\
	\inst{2} J. Stefan Institute, University of Ljubljana,
	         Jamova 39, 1000 Ljubljana, Slovenia}

\pacs{30.-v}{Liquid crystals}
\pacs{41.+e}{Polymers, elastomers and plastics}

\begin{document}

\maketitle

\begin{abstract}
We investigate the pseudo-Casimir force in a slab of material composed
of nematically ordered long polymers.  We write the total mesoscopic
energy together with the constraint connecting the local density and
director fluctuations and evaluate the corresponding fluctuation free
energy by standard methods.  It leads to a pseudo-Casimir force of a
different type than in the case of standard, short molecule nematic.
We investigate its separation dependence and its magnitude and
explicitly derive the relevant limiting cases.
\end{abstract}

Casimir effect is due to constrained fluctuations in media with long
ranged correlations \cite{Trunov}.  The physical nature of the medium
is not particularly important.  Though the standard Casimir effect has
been introduced for constrained electromagnetic field fluctuations
\cite{Casimir} it has been realized that other systems with long-range
correlations exhibit a similar type of fluctuation driven
interactions.  Most notably critical fluids \cite{KrechK}, smectic
manifolds \cite{Li} and liquid crystals \cite{Ajdari}, all of them
being prime examples of correlated fluids, give rise to a
pseudo-Casimir effect which comes about through constrained thermal
(as opposed to quantum) fluctuations of order parameters.  The nature
of these order parameters of course depends on the system under study
but they all exhibit massless fluctuation spectra that eventually lead
to long-ranged fluctuation interactions.

Delimiting ourselves to the case of liquid crystals, the
pseudo-Casimir effect has been researched in nematic, smectic and
columnar uniformly ordered systems \cite{Ajdari}.  In the case of
inhomogeneous or frustrated order as in the case of the hybrid-aligned
cell characterized by opposing surface fields or in the
Fr\'eedericksz cell where frustration arises from competing bulk and
surface fields \cite{Ziherl}, the interaction induced by director
fluctuations is enhanced substantially by frustration, the enhancement
being progressively stronger as the system approaches the transition
from uniform to distorted structure.

There seems to be some experimental evidence that backs up these
theoretical predictions \cite{Swanson,Musevic} but there still lacks a
definitive proof of the observability of the pseudo-Casimir
interaction.  It appears however that in the case of spinodal
dewetting of 5CB on a silicon wafer \cite{Vandenbrouck99} the
pseudo-Casimir interaction is essential in giving a consistent
interpretation of experimental data, implying that this experiment can
be regarded as the first observation of the pseudo--Casimir effect in
liquid crystals \cite{Ziherlprl}.

In this contribution we shall investigate the pseudo-Casimir effect in
the realm of confined polymers.  As a first case study we will take a
nematic polymeric liquid crystal such as one can observe in stiff
polyelectrolytes above the isotropic - nematic transition, confined
between two apposed planar surfaces.  A physical realization of this
case would be a nematically ordered DNA confined between two surfaces
or simply cut to form a slab of macroscopically oriented sample
\cite{Strey}.  We believe that this case is particularly instructive
since we will be able to connect different limiting results with the
existing calculations for ordinary, {\sl i.e.} short chain, nematics
\cite{Ajdari}.  It will be shown that the polymeric nature of the
nematogens gives rise to additional features of the pseudo-Casimir
effect that distinguish it qualitatively from results derived in the
case of short nematogens.  Also we will show that in the case of
nematic polymers the pseudo-Casimir interaction depends on the
equation of state \cite{Strey} of the polymer nematic through its
compressibility.

The elastic deformation of an ordinary short molecule nematic
\cite{deGennesProst} with long range orientational order along the
axis z, can be described with an average director ${\bf n}({\bf r})$,
with small fluctuations in the $(x,y)$ directions: ${\bf n}({\bf r})
\simeq (1,\delta n_{x}, \delta n_{y})$.  In this case, if the splay,
twist and bend elastic constants are the Frank constants $K_1$, $K_2$
and $K_3$ one obtains for the mesoscopic elastic hamiltonian
\begin{equation}
{\cal H}_{\bf n}=\ul12\int d^2{\bf r}_{\perp} dz \le [ K_1\le(
\bnabla_{\perp}\cdot \delta{\bf
n}\ri)^2 + K_2\le( \bnabla_{\perp} \times \delta{\bf n}\ri)^2 +
K_3\le( \partial_{z}\delta{\bf n}\ri )^2 \ri ].
\end{equation}
Besides fluctuations in nematic director one should also consider the
fluctuations in local density of the molecules.  For short nematogens
the director and density fluctuations are decoupled and one need not
consider this part of the mesoscopic free energy explicitly.  For
long, polymer nematics the situation is altogether different.  In this
case we have to consider the part of the free energy due to
nonhomogeneous density fluctuations $\rho ({\bf r}) \simeq \rho_{0} +
\delta \rho({\bf r})$ in the Ornstein-Zernicke form
\begin{equation}
{\cal H}_{\rho}=\ul12\int d^2{\bf r}_{\perp}dz \le [B(\delta\rho)^2 +
B\xi^2\le (\bnabla\delta\rho\ri )^2\ri ],
\end{equation}
where $B$ is the osmotic compressibility modulus and $\xi$ is the
density correlation length. Because polymers are here considered to be
infinitely long one also has to consider the fact that director
inhomogeneities can relax only if accompanied by the simultaneous
density relaxation \cite{coupled}. This leads to the constraint
\begin{equation}
0 = \bnabla \left( \rho {\bf n}\right) \simeq \partial_{z} \delta \rho +
\rho_{0} \left( \bnabla_{\perp}\cdot \delta{\bf n} \right).
\label{con}
\end{equation}
Adding all the components of the mesoscopic free energy and taking
the constraint Eq. \ref{con} into account via a Lagrange multiplier
we obtain
\begin{eqnarray}
{\cal H} = \ul12\int d^2{\bf r}_{\perp}dz && \left[  K_1\le(
\bnabla_{\perp}\cdot \delta{\bf
n}\ri)^2 + K_2\le( \bnabla_{\perp} \times \delta{\bf n}\ri)^2 +
K_3\le( \partial_{z}\delta{\bf n}\ri )^2 + \right. \nonumber\\
& & + B(\delta\rho)^2 + B\xi^2\le (\bnabla\delta\rho\ri )^2 + \left.
C \left( \partial_{z} \delta \rho +
\rho_{0} \left( \bnabla_{\perp}\cdot \delta{\bf n} \right) \right)^{2} \right].
\label{eq:0}
\end{eqnarray}

Writing $\delta{\bf n}$ in the Helmholtz ansatz $ \delta{\bf n} =
\delta{\bf n}_{\parallel}  + \delta{\bf n}_{\perp} $ we realize that
the tangential part of the hamiltonian is decoupled from
the constraint and its contribution to the fluctuational free energy
should be the same as for short nematics \cite{Ajdari}.

Obviously in the case of polymer nematics the longitudinal director
and the density field fluctuations are coupled.  The fluctuational
free energy is now obtained via
\begin{equation}
      {\cal F} = - kT ~{\rm ln}{\cal Z} = - kT ~{\rm
ln}{\int_{\delta{\bf n}(\partial )=0}{\cal
      D}\delta{\bf n} ~\int_{\delta\rho (\partial )=0 }  ~ {\cal
      D}\delta\rho~ \exp{(-\beta {\cal H})}},
      \label{eq:1}
\end{equation}
where we have assumed that at the boundary of the sample the
fluctuations in density and director field are quenched ({\sl i.e.}
$\delta{\bf n}(\partial )=0, \delta\rho (\partial )=0$). In principle
one could also introduce the anchoring and tension energy of the
bounding surfaces which would not bring any new qualitative features
into our discussion but would make the computations less transparent.

We now analyze the fluctuational free energy of a slab of nematic
material bounded at $z = \pm L/2$ where the average director is
aligned with the $z$ axis. In the case of short nematogens ($C =
0$) the free energy decouples and we obtain two contributions. A long
range pseudo-Casimir free energy due to director fluctuations and a short
range term due to screened density fluctuations. Thus
\begin{equation}
      {\cal F}(C=0)  = - \frac{kTS}{16 \pi} \le (
      \frac{K_{3} }{K_{1} } + \frac{K_{2} }{K_{1} }\ri )
      \frac{\zeta (3)}{L^{2} } -  \frac{kT}{16 \pi}
      \frac{\epsilon (L/\xi )}{L^{2} },
      \label{eq: }
\end{equation}
where $S$ is the area of the surface, $\zeta (n)$ the Riemann's Zeta
function and
\begin{equation}
\epsilon (L/\xi ) = \int_{2L/\xi}^{\infty}  u~ du
~{\rm ln} \le ( 1-~e^{-u}  \ri ).
\end{equation}

Since $L/\xi \gg 1$ and the function $\epsilon (x)$ has an
exponential behavior for large values of the argument the density
fluctuations make a negligible contribution to the total free
energy.  Our result thus evidently reduces to the one obtained by
Ajdari {\sl et al.} \cite{Ajdari}.

In the case that the director and the density fluctuations are
coupled the calculation of the functional integral Eq. \ref{eq:1}
becomes more complicated. First of all we transform the hamiltonian
Eq. \ref{eq:0} by introducing
\begin{equation}
\delta\rho = \rho_{0} \bnabla_{\perp} \cdot {\bf a}+ f \quad\quad , \quad\quad
\delta {\bf{n}}=-\partial_{z} {\bf a}= -\dot{\bf a}.
\label{const}
\end{equation}

The coupling term in this case reduces to

\begin{equation}
C\left ( \partial_{z} {\rho} + \rho_{0} \bnabla_{\perp}\cdot
\delta{\bf n} \right )^2 \to
C\rho_{0}^{2} \left ( \partial_{z} {f}\right )^2 = C'\dot{f}^2
\end{equation}

and the hamiltonian is

\begin{eqnarray}
{\cal H}=\ul12\int d^2{\bf r}_{\perp}dz\Biggl[
&K_3 & \ddot{\bf a}^2 + K_1 \le (\bnabla_{\perp}\dot{\bf{a}}\ri )^2
+ B\le (\bnabla_{\perp}{\bf{a}}+ f \ri )^2
+B\xi^2 \le (\bnabla_{\perp} (\bnabla_{\perp}{\bf{a}} + f) \ri
)^2+ \nonumber\\
&+& B\xi^2\le (\bnabla_{\perp}\dot{\bf{a}}+ \dot{f} \ri )^2
+C'\dot{f}^2 \Biggr ]\; .
\end{eqnarray}

Clearly in the limit of strictly enforced density-director coupling
$C \longrightarrow \infty$ and as a consequence $f$ becomes
independent of $z$, thus $\dot{f} = 0$. In this case we can introduce
the following linear transformation
\begin{equation}
{\bf a} \longrightarrow {\bf a} + \bnabla_{\perp}  A , \quad\quad
\nabla_{\perp} ^{2}A = -f ,
\end{equation}
obtaining finally for the hamiltonian in the limit of strong coupling

\begin{equation}
{\cal H}\!=\!\ul12\int d^2{\bf r}_{\perp}dz\Biggl[
K_3 \ddot{\bf{a}}^2\!+\!K_1 \le (\bnabla_{\perp}\!\cdot\!\dot{\bf{a}}\!\ri )^2
\!+\!B\le (\bnabla_{\perp}\!\cdot\!\bf{a}\!\ri )^2
\!+\!B\xi^2 \le (\bnabla_{\perp} (\bnabla_{\perp}\!\cdot\!\bf{a})\!\ri )^2\!+\!
B\xi^2\le (\bnabla_{\perp}\!\cdot\dot{\bf{a}} \ri )^2 \Biggr ] .
\end{equation}

Since ${\bf a}$ is a longitudinal vector (Eq. \ref{const}) one
obtains in the Fourier space

\begin{eqnarray}
{\cal H} &=& \ul12 \sum_{\bf Q}~ \int dz  \Biggl[
K_{3} \ddot{\bf{a}_{\parallel}}^2 + \le (K_1 + B\xi^2 \ri )
Q^{2}\dot{\bf{a}_{\parallel}}^2 + B \le ( 1 + \xi^{2} Q^{2} \ri )
Q^{2} \bf{a}_{\parallel}^2 \Biggr ]\; .  \nonumber\\
~
\end{eqnarray}

The mesoscopic hamiltonian obviously corresponds to a persistent
oscillator, {\sl i.e.} an oscillator with an additional fourth order
term in the derivatives.  Since the different Fourier components are
decoupled the evaluation of the functional integral proceeds
straightforward and can in fact be reduced to a Feynman integral
for second-derivative Lagrangian solved exactly by Kleinert
\cite{Kleinert,handbook}.  Taking into account the boundary conditions for the
fluctuating fields we obtain up to a multiplicative constant

\begin{eqnarray}
      {\cal Z} = \frac{(2\pi kT)^{-1} \sqrt{\omega_{1}\omega_{2}}~\vert
      \omega_{1}^{2} -
      \omega_{2}^{2}\vert  (\omega_{1}^{2} +\omega_{2}^{2})^{-1}  }{{\rm
      sinh}(\omega_{1}K_{3}^{-1/3}L){\rm sinh}(\omega_{2}K_{3}^{-1/3}L)
      - \frac{2\omega_{1}\omega_{2}}{(\omega_{1}^{2} +\omega_{2}^{2})
}({\rm cosh}(\omega_{1}K_{3}^{-1/3}L){\rm
      cosh}(\omega_{2}K_{3}^{-1/3}L) - 1) }, \nonumber\\
      ~
\end{eqnarray}
where
\begin{eqnarray}
      (\omega_{1}^{2} +\omega_{2}^{2}) &=& \ul12 \le (K_1 + B\xi^2 \ri )
      Q^{2} K_{3}^{-1/3} \nonumber\\
      (\omega_{1}\omega_{2})^{2} &=& B \le ( 1+ \xi^{2} Q^{2} \ri )
Q^{2}K_{3}^{1/3}.
      \label{eq:7}
\end{eqnarray}

Introducing now $\Omega_{\pm} = K_{3}^{-1/3} 
(\omega_{1}\pm\omega_{2})$ and with

\begin{equation}
\Omega^2_{\pm}(Q)=\frac{K_1 +B\xi^2 }{2K_3} Q
\Biggl [ Q \pm 4\Lambda\sqrt{1+\xi^2 Q^2}\Biggr ]\; .
\end{equation}

where

\begin{equation}
\Lambda = \frac{\sqrt{BK_3}}{K_1+B\xi^2}
\end{equation}

we obtain for the regularized fluctuation free energy per unit area,
where the formally divergent bulk and
surface free energy terms have already been subtracted
\begin{equation}
      {\cal F} = \frac{kTS}{4\pi} \int_{0}^{\infty} QdQ ~{\rm ln} \le ( 1 - 2
      \frac{\Omega_{+}^{2}}{\Omega_{-}^{2}}
e^{-\Omega_{+}L} \le [ {\rm
cosh}(\Omega_{-}L) -
      \frac{\Omega_{+}^{2} - \Omega_{-}^{2}}{\Omega_{+}^{2}} \ri ] +
      e^{-2\Omega_{+}L}\ri ).
      \label{eq:8}
\end{equation}

In the above equations $\Omega_{-}^2$ can become negative for sensible
values of the correlation length $\xi$.  In this case the ${\rm
cosh}{\Omega_{-} L}$ has to be replaced by $\cos{\vert\Omega_{-}\vert
L}$ but the integral contains no dangerous divergencies and can be
evaluated straightforwardly.  By introducing
\begin{equation}
{\cal F}^{*}  = \frac{kTS}{4\pi} \Lambda^{2} \quad {\rm and} \quad L^{*}  =
\sqrt{\frac{2(K_1+B\xi^2)}{B}}
\end{equation}
we realize that the dimensionless free energy ${\cal F}/{\cal F}^{*}$
obtained from Eq.  \ref{eq:8} depends only on dimensionless separation
$L/L^{*}$ and the dimensionless coupling constant $\gamma = \Lambda
\xi$. This coupling constant basically represents the ratio between
the density correlation length $\xi$ and the polymer nematic
correlation length perpendicular to the average director
$\Lambda^{-1}$ \cite{Strey}.

\begin{figure}[ht]
\begin{center}
      \epsfig{file=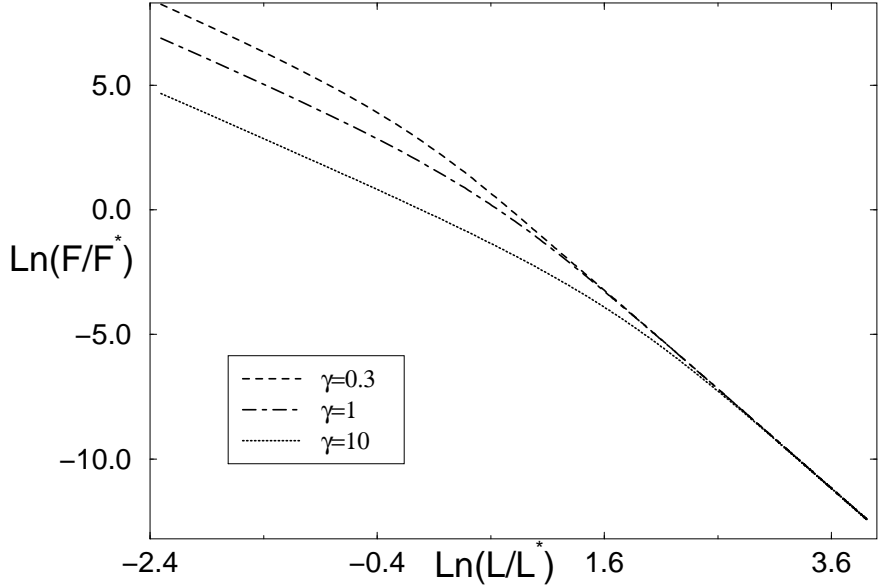, width=8cm, angle=-90}
\end{center}
\caption{ The dependence of the dimensionless free energy $\frac{\cal
F}{\cal F}^{*}$ on the dimensionless separation $\frac{L}{L^{*} }$
obtained from Eq.  \ref{eq:8} for different values of the coupling
constant $\gamma$.  The two scaling regimes at small and large
$\frac{L}{L^{*} }$ are clearly visible.}
\label{fig1}
\end{figure}

\begin{figure}[h]
\begin{center}
      \epsfig{file=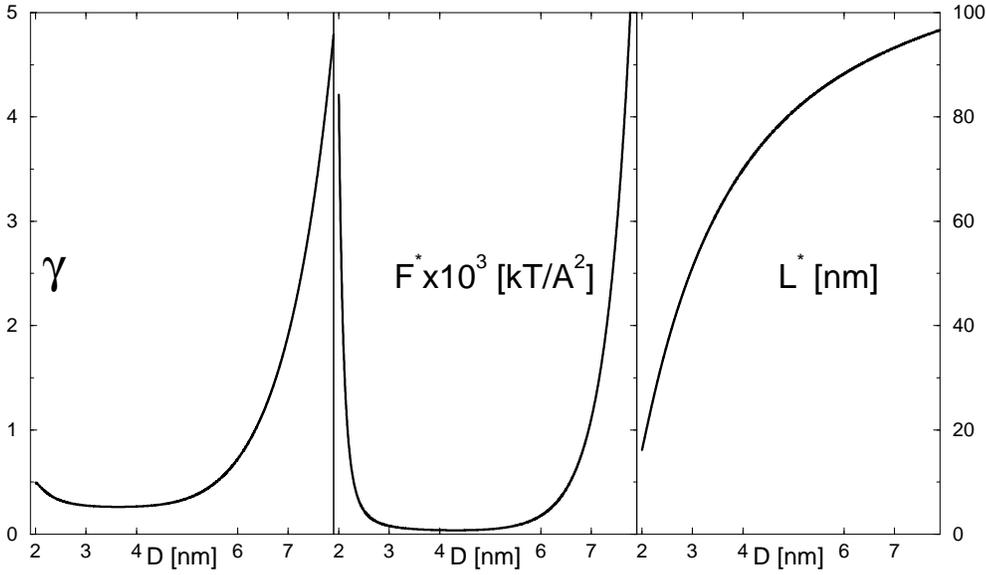, width=8cm, angle=-90}
\end{center}
\caption{ The dependence of the magnitude of the pseudo-Casimir
interaction ${\cal F}^{*} $, the characteristic length $L^{*}$ and the
coupling constant $\gamma$ on the average separation between the
polymers obtained from \cite{Strey} in the case of 0.5 M bathing ionic
solution.  Note that $\gamma$ and ${\cal F}^{*}$ have the same scale.
$L^{*}$ saturates for small values of DNA density (large $D$) at
about $140~ \AA$.}
\label{fig12}
\end{figure}

In order to estimate the magnitude of the pseudo-Casimir forces in
confined nematic polymers we now have to connect the macroscopic
elastic constants $K_{1} , K_{3} , B$ and the correlation length $\xi$
with the microscopic parameters of the systems.  This is in general a
very difficult undertaking.  For DNA this type of analysis has been
performed in \cite{Strey , Strey2} and  we just quote these results.

If the interaction potential between the segments of the polymers is
$U(D)$, where $D$ is the average spacing between the molecules
perpendicular to the average director and the persistence length of
the polymers is ${\cal L}_{P} $, one has \cite{Strey}
\begin{eqnarray}
     K_{1}  = K_{2}  &\simeq & U(D)/D \nonumber\\
     K_{3}  &\simeq & kT~{\cal L}_{P}~\rho_{\perp} + U(D)/D\nonumber\\
     B &\simeq & \frac{\sqrt{3}}{4} \le (
\frac{\partial^{2} }{\partial D^{2} } -
     \frac{1}{D}\frac{\partial}{\partial D} \ri ) U(D)/{\cal L}_{P}.
\end{eqnarray}
One should realize here that since DNA is a polyelectrolyte the
interaction potential $U(D)$ depends also on the ionic strength of
the bathing solution.

This scaling of elastic constants leads to the conclusion that in Eqs.
\ref{eq:9} and \ref{eq:10} we have $\frac{K_{2} }{K_{1} } \sim 1$ and
$\frac{K_{3} }{K_{1} } \gg 1$.  Thus the larger the $D$, {\sl i.e.} the
smaller the density of the polymers, the larger this ratio becomes.
The magnitude of the pseudo-Casimir interaction thus depends
fundamentally on the equation of state of the nematic polymers through
the osmotic compressibility $B$ and the density perpendicular to the
average nematic director, $\rho_{\perp} $.

Fig.  1 presents a plot of dimensionless free energy as a function of
dimensionless separation. The curves are plotted for various values of
the coupling constant $\gamma$ which is plotted as a function of the
average spacing between polymers on Fig. 2 for the case of DNA.  One
can clearly discern two regions with approximately $L^{-2}$ and
$L^{-4}$ behavior separated by a characteristic spacing $L_{0} =
C(\gamma) L^{*} $, where $C(\gamma)$ is a numerical factor depending
on the coupling constant that can be read off Fig. 1. The numerical
evidence shows that the caracteristic spacing $L_0$ grows
approximately linearly with the average separation between the
polymers $D$, ranging from about 15 to 20 nm at $D=2$ nm to around 150
nm at $D=8$ nm, which is between one and two correlation lengths $\xi$.

Fig.  2 shows the dependence of the magnitude of the pseudo-Casimir
interaction ${\cal F}^{*} $, the characteristic length $L^{*} $ and
the coupling constant $\gamma$ on the average separation between the
polymers obtained from the measured equation of state for DNA
in the case of $0.5 M$ bathing electrolyte solution (for details see
\cite{Strey} ).

One would of course like to have also approximate analytical formulae
for the pseudo-Casimir interaction Eq.  \ref{eq:8}.  There are only two
limiting cases where the integral in Eq.  \ref{eq:8} can be evaluated
analytically, the large separation and the small separation regime.
First of all for small $L$ the dominant behavior of the integral is
obtained at large $Q$ leading in this case to
\begin{equation}
      {\cal F}(L\ll L_{0})=-~\frac{kTS}{16 \pi L^2} \le (
      \frac{K_3}{K_1 \le ( 1+4\Lambda\xi\ri )\cdot
      \le ( 1+\frac{B}{K_1}\xi^{2}\ri ) } +
      \frac{K_2}{K_1} \ri )\zeta (3).
\label{eq:9}
\end{equation}

The other analytically tractable limit is in the case of large $L$
where the dominant contribution to the integral comes from small $Q$
behavior of the integrand
\begin{equation}
      {\cal F}(L \gg L_{0} ) = - ~\frac{kTS}{16 \pi} ~\le [
      \frac{2K_{3}\zeta (5)}{BL^{4} }
      +\frac{K_{2}\zeta (3)}{K_{1} L^2} \ri ],
\label{eq:10}
\end{equation}
where the length scale $L_{0}$ can be read off the graph on Fig.  1
for various values of $\gamma$ (see above).  Thus for small $L$ the
confined polymer nematic behaves basically as a standard nematic
\cite{Ajdari} with a renormalized magnitude of the fluctuation
interaction. For large $L$ the dependence is altogether different,
deviating essentially from the standard nematic case. The crossover
between the two regimes moves to larger $L$ as the polymers become
less dense (see Fig. 2) but eventually saturates for small polymer
densities.

The pseudo-Casimir interaction for long nematic polymers thus differs
qualitatively from the one obtained in the case of short nematics
\cite{Ajdari}.  Though it still decays algebraically with $L$, and
thus qualifies as a long range force, it decays faster ($L^{-4} $)
then in the case of short nematics ($L^{-2} $).  Nevertheless its
absolute magnitude, depending on the equation of state of the nematic
polymer and the regime of spacings $L$, can become comparable to and
even larger than in the case of short nematics with the same Frank
constants.  The $L^{-4}$ behavior stems essentially from the polymer
stiffness and should thus be a salient feature of the pseudo-Casimir
interactions whenever the mesoscopic hamiltonian contains the squares
of higher then the first derivatives in the order parameter.  These
type of systems will be studied in our future work.

\acknowledgments
We would like to thank P. Ziherl for his valuable comments on an
earlier version of the MS.

\end{document}